\documentclass[prl,aps,showpacs,twocolumn,floatfix]{revtex4-2}
\usepackage{graphicx}
\usepackage{bm}
\usepackage{amsmath}
\usepackage{amssymb}

\begin{document}

\title{Entanglement Entropy in Non Relativistic Collisions}

\author{B. G. Giraud}
\affiliation{Institut de Physique Th\'eorique,
Centre d'Etudes Saclay, 91191 Gif-sur-Yvette, France, Bertrand.Giraud@cea.fr}
\author{R. Peschanski}
\affiliation{Institut de Physique Th\'eorique,
Centre d'Etudes Saclay, 91191 Gif-sur-Yvette, France, Robi.Peschanski@cea.fr}

\begin{abstract}
We study numerically the build-up of entanglement entropy between two
colliding, non relativistic particles, that are driven by the plain
Schr\"odinger equation containing local, finite range potentials, and whose
initial state is the product of two independent wave packets.
Different situations are considered, ranging from
a tunnel and mirror effect at low energy in front of a barrier, to higher
energies, and also attraction rather than repulsion. While the calculated
wave functions and density matrices significantly depend on the physical
situations described by our models, the build-up trends and values in time of
entanglement entropy look surprisingly alike one another.



\end{abstract}

\date{\today}

\maketitle

\section{I Introduction}
In a recent paper \cite{robi} the relativistic  ${\bf{S}}$-matrix theory of
collisions was shown to allow a measure of entanglement between either the
elastic or inelastic collison of two elementary particles. However, since the
${\bf{S}}$-matrix, by definition involves asymptotic times only, a question
was left open, namely how does entanglement evolve at finite times?
The present work attempts to answer the question, by means of a non 
relativistic approach, where one can use time-dependent wave functions. Given 
two particles, labelled $1$ and $2$, we generate numerically soluble models 
providing wave functions $\psi(1,2;t)$ and the corresponding density matrices 
$\rho_{12}(t)$, $\rho_1(t)$ and $\rho_2(t)$ and corresponding entropies 
$S_{12}(t)$, $S_1(t)$ and $S_2(t)$. The time evolution of the entanglement 
entropy, ${\cal I}=S_1+S_2-S_{12}$, can then be calculated and observed.

The effect of scattering on the entanglement of two nonrelativistic colliding
particles has been studied in general in Ref.\cite{mishima}. In
the one-dimensional case, dedicated studies have been performed either using a
general framework for the potential \cite{Harshman:2007omw}, wave packets for
the particles, \cite{law}, or an analytic example with a hard-core potential
\cite{jantzig}. The novelty of our approach is the full numerical computation
of the  entanglement entropy for various cases of potentials and its comparison
with the relativistic case of Ref.\cite{robi}.

Section II of this work describes our formalism. Section III presents initial
conditons without entanglement, a useful simplification that does not bias
our arguments. Section IV describes our numerically soluble models and their
results. Section V discusses the existence of an asypmtotic entanglement.
Section VI is our conclusion. An Appendix states two theorems which facilitate
understanding the time evolution of entropies.

\section{II Basic formalism in an interacting two-body problem}

We use a system of units where $\hbar=1$. Consider in dimension $d$, ($d=3$
for instance,) two non-relativistic, spinless particles, of equal mass $m$,
$m=1$, hence relative mass $\frac{1}{2}$, with coordinates
$\vec r_i \equiv \{x_i,y_i,z_i\},\ i=1,2$, and their associated momenta
$\vec p_i$. Their center of mass (cm) degrees of freedom read,
$\vec R \equiv \{X,Y,Z\} = \frac{1}{2} \left(\vec r_1 + \vec r_2 \right)$ 
and $\vec P = \vec p_1+\vec p_2$. Their relative motion (rel) degrees of freedom
read $\vec r \equiv \{x,y,z\} = \vec r_1 -\vec r_2$, with 
$\vec p\equiv \{p_x,p_y,p_z\}=\frac{1}{2} \left(\vec p_1 - \vec p_2 \right)$.
For the sake of simplicity, we set this particle pair as driven by a relative,
Galilean invariant, scalar and local potential, $V(r),\ r=| \vec r |$. For
instance, $v$ can be a finite range repulsion or attraction,
$V(r)=\lambda e^{-\mu r^2}$.

Their Hamiltonian reads, $H=(p_1^2+p_2^2)/2+V(r)$. Notice that this $H$ is
also the sum of two independent Hamiltonians, $H=\left(p^2+V(r)\right)+P^2/4$,
$H_{rel}=p^2+V(r)$ and $H_{cm}=P^2/4$. The whole wave function can be chosen
as a product $\psi_{cm}(\vec R;t) \psi_{rel}(\vec r;t).$ Hence, the cm
is a passive degree of freedom, that factorizes out. Accordingly, its state
can be chosen at first as some wave packet $\psi_{cm}(\vec R;0)$. Let 
$\beta$ be a reference length for single particle states. Then we choose this 
$\psi_{cm}$ as a spherical Gaussian,
\begin{equation}
\psi_{cm}(\vec R;0)=
(2/\pi)^{\frac{d}{4}}\, \beta^{\, -\frac{d}{2}}\, \exp[-R^2/\beta^2],\  R=| \vec R |.
\label{pscm}
\end{equation}
At finite times, it evolves under spreading and phase oscillations only,
\begin{equation}
\psi_{cm}(\vec R;t)=
(2/\pi)^{\frac{d}{4}}\, \beta^{\, \frac{d}{2}}\, (\beta^2 +i t)^{-\frac{d}{2}}\,
\exp[-R^2/(\beta^2+i t)].
\label{pscmt}
\end{equation}

We define entanglement as follows: whether one handles a time independent
theory with a wave function $\psi(\vec r_1,\vec r_2)$ or a time dependent one
with a wave function $\psi(\vec r_1,\vec r_2;t)$, the question to be raised is
whether there exists single particle wave functions $\varphi_1$, $\varphi_2$,
such that $\psi(1,2) = \varphi_1(1)\, \varphi_2(2)$.
More precisely, let $\psi$ evolve in time and let the density matrix (dm) of 
the system be, at all times, the pure dyadic, 
$\rho_{12}=| \psi \rangle \langle \psi |$, and then note that the
corresponding entropy, $-Tr ( \rho_{12} \log{\rho_{12}} )$, trivially vanishes
at all times. Let $\rho_{12}$ be represented as a function
$\rho_{12}(\vec r_1,\vec r_2,\vec r_1^{\ \prime},\vec r_2^{\ \prime})$, or,
in a shorter notation, as $\rho_{12}(1,2,1',2')$.
The partial traces, $\rho_1=Tr_2\,\rho_{12}$ and $\rho_2=Tr_1\,\rho_{12}$, define
reduced dm's for particles 1 and 2, respectively. Then the combination,
\begin{equation}
{\cal I}=
Tr(\rho_{12} \log{\rho_{12}})-Tr(\rho_1 \log{\rho_1})-Tr(\rho_2 \log{\rho_2}),
\label{defini}
\end{equation}
a semipostive quantity, measures the entanglement, namely how much the product,
$\rho_{12}(1,2,1',2')=\psi(1,2) \psi^*(1',2')$, differs from the
product $\rho_1(1,1')\, \rho_2(2,2')$.

\section{III Initial conditions without entanglement} 

For a vanishing quantity ${\cal I}$ at beginning, set single particle
states,
$\varphi_1(\vec r_1;0)=\pi^{-\frac{d}{4}} \beta^{-\frac{d}{2}}
\exp[ i\vec k.\vec r_1 - ( \vec r_1 + \vec \sigma )^2/(2 \beta^2)]$,
and
$\varphi_2(\vec r_2;0)=\pi^{-\frac{d}{4}} \beta^{-\frac{d}{2}}
\exp[-i\vec k.\vec r_2 - ( \vec r_2 - \vec \sigma )^2/(2 \beta^2)]$,
with a vector $\vec \sigma$ to separate the particles from each other and a
vector $\vec k$ to boost them against each other. Here, $\beta$ is the same
as that length used to define $\psi_{cm}$ in the previous section.
Choose as an initial wave function for the pair of particles the product,
$\psi(1,2;0)=\varphi_1 \varphi_2$. This induces a vanishing two-body entropy,
that remains vanishing under time evolution, as explained in the Appendix.
It is then also trivial that this initial product, $\psi(1,2;0)$, 
factorizes as the cm wave function $\psi_{cm}(\vec R;0)$, that packet 
defined in the previous section, multiplied by a relative wave function,
$\psi_{rel}(\vec r;0)=
(2 \pi)^{-\frac{d}{4}} \beta^{-\frac{d}{2}} \exp[i\vec k.\vec r]
\exp[-(\vec r+2 \vec \sigma)^2/(4 \beta^2)]$.
It is again trivial that, still at $t=0$, with 
$\rho_{12} = | \psi \rangle \langle \psi |$,
then $\rho_1=| \varphi_1 \rangle \langle \varphi_1 |$ and 
$\rho_2=| \varphi_2 \rangle \langle \varphi_2 |$. Accordingly,
$Tr \rho_{12}=Tr \rho_1=Tr \rho_2=1$, and furthermore, again when $t=0$,
$Tr \rho_{12} \log{\rho_{12}}=Tr \rho_1 \log{\rho_1}=Tr \rho_2 \log{\rho_2}=0$
and thus, initially, ${\cal I}=0$. 

\subsection{The case when $V=0$}

The case, $V=0$, makes a trivial situation. Since $H$ reduces to
$p_1^2/2 + p_2^2/2$, the particle states $\varphi_1,\, \varphi_2$ evolve
independently, with just spreading and phase oscillations. All density
operators remain dyadics, all entropies vanish at all times, hence ${\cal I}$
remains vanishing. It is not surprising that finite entanglement demands an
activted interaction. See the Appendix for more details.
  
\section{IV Numerical illustrations in 1d}

For the sake of simplicity, we now consider {\it  1d} toy models only,
because higher dimensions, while numerically tractable, demand heavy
calculations. Now $\vec r$ has only one component $x$, and the other
degrees of freedom also reduce to a single component. The initial state reads, 
$\psi(x,X;0)=\psi_{rel}(x;0) \psi_{cm}(X;0)$. The cm evolution reads,
\begin{equation}
\psi_{cm}(X;t) = (2/\pi)^{\frac{1}{4}} \sqrt{\frac{\beta}{\beta^2+i t}}
\exp\left[-\frac{X^2}{\beta^2+i t}\right] .
\label{cmevlv}
\end{equation}

The oncoming illustations explore four physical cases, namely 1- a repulsive
barrier with a coexistence of tunnel and mirror effects, 2- the same barrier
when the relative particle has an energy equal to twice its height, 3- still
the same barrier when the energy equals its height and 4- the same energy
in front of an attractive pocket of shape and depth equal to the shape and
height of the previous barriers.

In all cases we use numerical solutions of the equation, 
$i d\psi_{rel}/dt=H_{rel} \psi_{rel}$, to calculate the time evolution of
$\psi_{rel}$ in the $x$-representation. Then, when needed, we can use the
corresponding density matrix,
$\rho_{rel}=| \psi_{rel} \rangle \langle \psi_{rel} |$. 
Simultaneously, the Gaussian 
$\psi_{cm}(X;t)$ listed just above, exhibits spreading, and phase
oscillations. All told, single particle density operators will be
obtained from partial traces, with $x_2=x'_2$ or $x_1=x'_1$, of the single
particle coordinate representation of the full dm, $\rho_{12}=
\psi_{cm}[(x_1+x_2)/2;t]\, \psi_{rel}(x_1-x_2;t)\, \psi_{cm}^*[(x_1^{\, \prime}+
x_2^{\, \prime})/2;t]\, \psi_{rel}^*(x_1^{\, \prime}-x_2^{\, \prime};t)$.
We set $\beta=1$. Then we set  $\sigma=2$; this induces a
sufficient separation between the intitial particle states. 

\subsection{IV.1 Case when particles repel each other and a tunnel effect
occurs}
\begin{figure}
\scalebox{0.50}{\includegraphics*{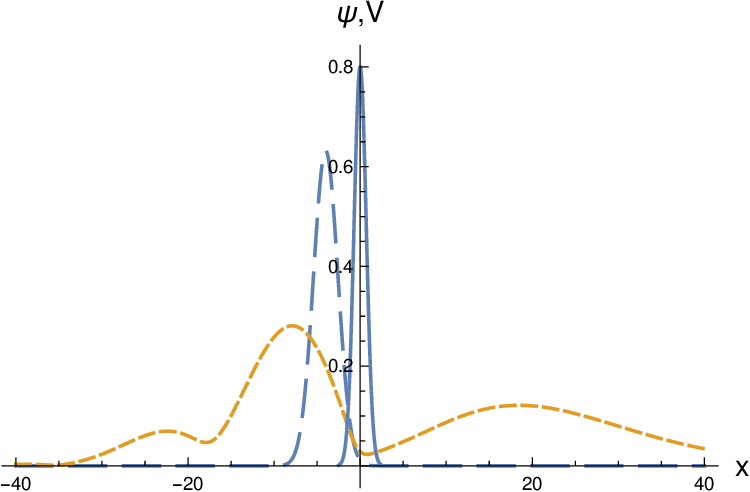}}
\caption{Evolution of $|\psi_{rel}(x;t)|$ if $k=.55$, $\sigma=2$,
and $V=e^{-r^2}$. Full line, potential, scaled down by factor $.8$. Long dash
line, $|\psi_{rel}|$ at $t=0$. Short dash line, $|\psi_{rel}|$ at $t=12$. Notice
both bounce and tunnel effects.}
\label{fintr1}
\end{figure}

The boost value, $k=0.55$, hence relative energy 0.3025, in presence of a
repulsion $V=e^{-r^2}$, induces both a significant tunnel effect and a
significant bouncing, as shown by Figure \ref{fintr1}, which displays the
modulus of $\psi_{rel}(x;t)$ at times $t=0$ and $t=12$. The figure also shows
$V$, scaled down by a coefficient $.8$ for graphical convenience.

\begin{figure}
\scalebox{0.50}{\includegraphics*{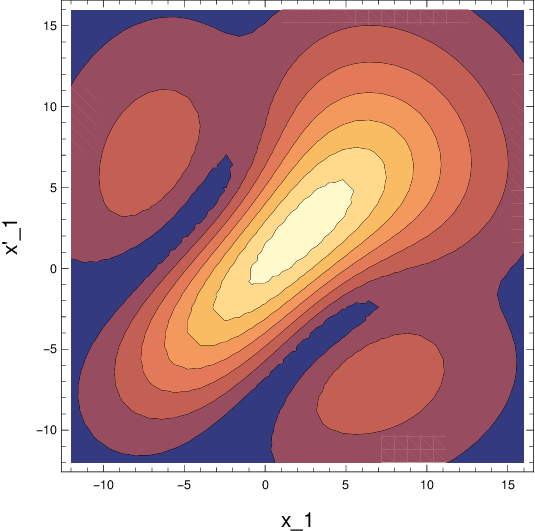}}
\caption{Same parameters as for Fig.\ref{fintr1}. Contour plot, at time t=8,
of the modulus of the density operator of particle 1.}
\label{fintr23}
\end{figure}
 
While at time $t=0$ the plots of the density operators of single particle 1
and 2 give trivial results, such as circular contour lines for their modulus
for instance, because we chose Gaussians as initial conditions, strong
deformations are generated by the time evolution. We looked at two examples,
the contour plots of $|\rho_1(x_1,x'_1,4)|$ and $|\rho_1(x_1,x'_1,8)|$.
The latter is shown as Fig.\ref{fintr23}. We verified
that, at each time, the density operator of the second particle results from a
trivial parity symmetry operation upon that of the first particle. Both
corresponding entropies will thus be the same. We find, at time $t=4$, the
result, $-Tr \rho_1 \log \rho_1 \approx 0.71$, hence, at that time, 
${\cal I} \approx 1.4$. At time $t=8$ we obtain ${\cal I} \approx 2.2$. Values
at further times seem to stabilize into an asymptotic limit.

\subsection{IV.2  Case with relative particle at a higher energy than a
barrier}
\begin{figure}
\scalebox{0.50}{\includegraphics*{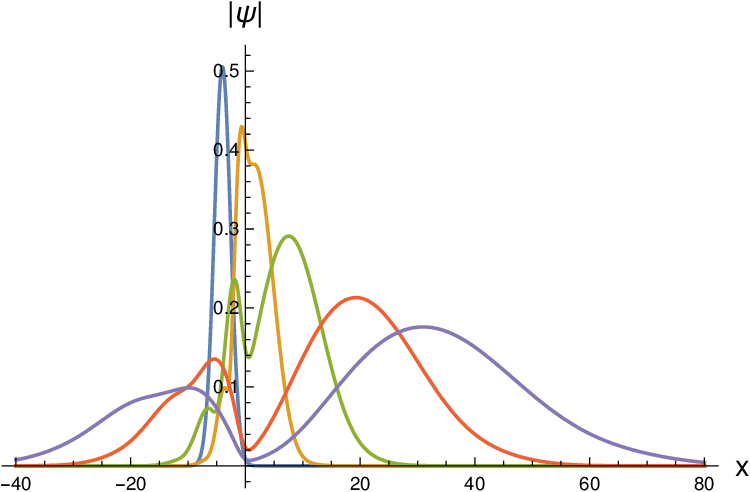}}
\caption{Same parameters as for Fig.\ref{fintr1}, but now $k=\sqrt{2}$.
Evolution of $|\psi_{rel}(x;t)|$. From left to right, peaks at $t=0,2,4,8,12$.
Peak for t=0 scaled down by factor $0.8$.}
\label{fintr34}
\end{figure}

With still $V=e^{-r^2}$, $\sigma=2$ and the same relative mass $1/2$, but now
$k=\sqrt{2}$, hence relative energy $2$, twice higher than the barrier,
we show in Figure \ref{fintr34} the modulus of $\psi_{rel}(x;t)$ at times
$t=0,2,4,8$ and $12$, their peaks ranging from left to right. For graphical
convenience, $|\psi_{rel}(r;0)|$ is multiplied by 0.8. Notice how, as soon
as $t \ge 4$, the peaks seem to evolve under simple translation and
spreading rules. We find ${\cal I} \approx 1.1$ and ${\cal I} \approx 1.3$ for
$t=4$ and $t=8$, respectively. At larger times, the entanglement $\cal{I}$
again seems to reach a stable, asymptotic value.

\subsection{IV.3 Case where the relative energy equals the height of the
barrier}
\begin{figure}
\scalebox{0.50}{\includegraphics*{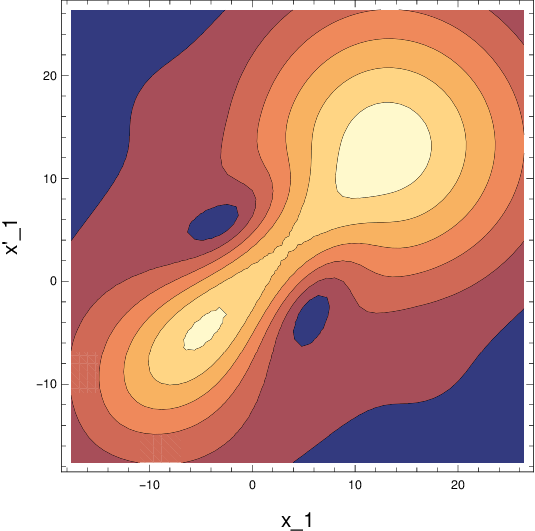}}
\caption{Same parameters as for Fig.\ref{fintr1}, but now $k=1$.
Contour plot of the density operator of particle $1$ at time $t=4$.}
\label{fig141}
\end{figure}

We retain the same barrier $V$ but here set $k=1$. Hence, the relative particle
is boosted at just the energy of the barrier top. We recover a situation where
similar amounts of transfer and reflection coexist. The result for the density
operator of particle $1$ at time $t=12$ is shown on Figure \ref{fig141}. 
Particle $2$ shows the same pattern under parity, naturally.
The entanglement entropies for times $t=4$ and $t=8$ are found to be worth
$\approx 1.6$ and $\approx 1.9$, respectively. There is again a hint of an
asymptotic behavior of $\cal{I}$.

\subsection{IV.4 Particles experiencing attraction}
\begin{figure}
\scalebox{0.50}{\includegraphics*{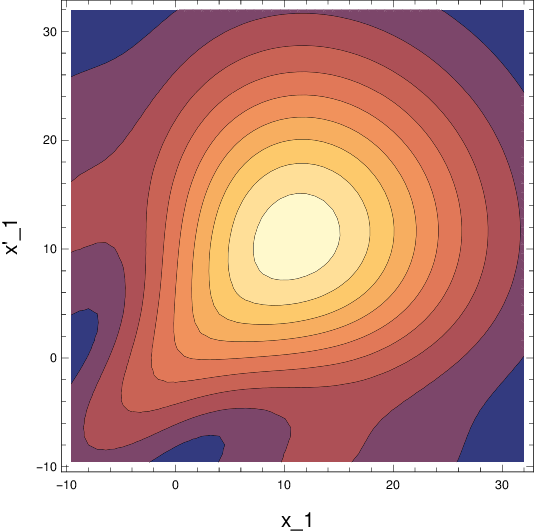}}
\caption{Still $\sigma=2$ and $k=1$, but now $V=-e^{-x^2}$. Contour plot of
the density operator of particle $1$ at time $t=12$.}
\label{fig14m1}
\end{figure}

Now $V$ is an attractive pocket, $V=-e^{-x^2},$ and we keep $\sigma=2$
and $k=1$. The contour plot of the density operator for particle $1$ is shown,
see Figure \ref{fig14m1} for $t=12$.
Times $t=4$ and $t=8$ provide for $\cal{I}$ the approximate values $1.0$
and $1.2$, respectively. Once more, values at further times seem to be
compatible with an asymptotic stability of $\cal{I}$, see the next section.

\subsection{IV.5 Preliminary conclusion}
\begin{figure}
\scalebox{0.50}{\includegraphics*{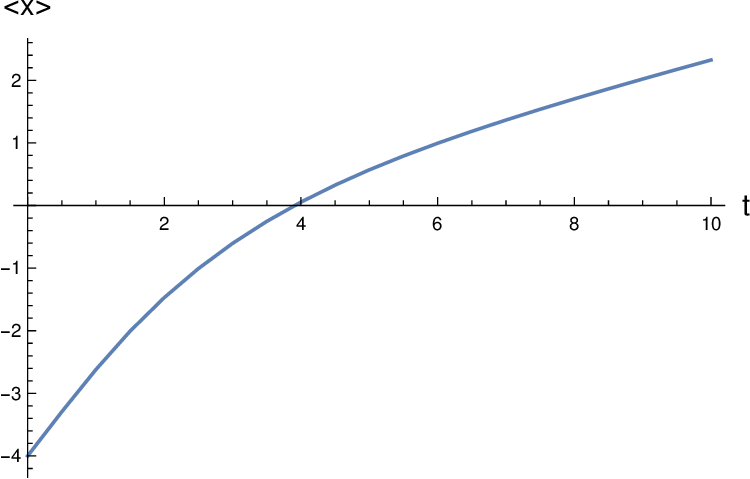}}
\caption{Evolution in time of the average separation $\langle x \rangle$ for
$V=e^{-x^2}$, $\sigma=2$ and $k=\sqrt{2}$.}
\label{separ}
\end{figure}

\begin{figure}
\scalebox{0.50}{\includegraphics*{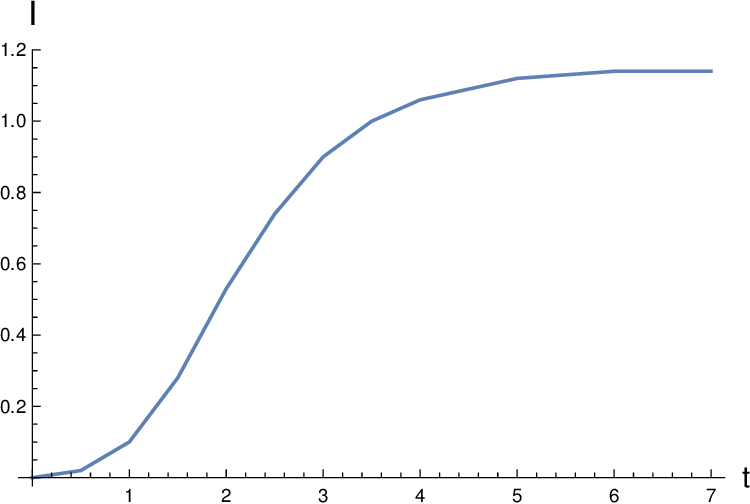}}
\caption{Same  parameters as for Fig. \ref{separ}. Evolution in time of the
entanglement entropy ${\cal I}$.}
\label{trica}
\end{figure}

It turns out that, in all four cases, ${\cal I}(t)$ shows similar behaviors
and orders of magnitude. As a typical example, we show in Figure \ref{separ}
and Figure \ref{trica} the behaviors in time of $\langle x \rangle$  and
${\cal I}$, respectively, for the second of our cases above,
namely that where $k=\sqrt{2}$ with the barrier, $V=e^{-x^2}$.

\section{V Entanglement at large times}

\begin{figure}
\scalebox{0.50}{\includegraphics*{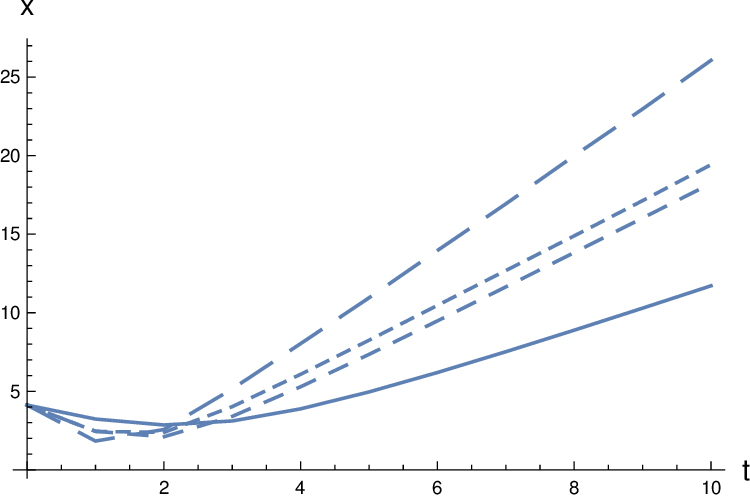}}
\caption{Full line, evolution in time of $\sqrt{\langle x^2 \rangle}$ for our
first toy illustration. Long dashes, moderate ones, small ones, the same
evolution for our second, third and fourth numerical illustrations.}
\label{figxbar}
\end{figure}

It is intuitive, when $V$ is a finite range potential, that this potential
becomes switched off when time increases. Indeed, while the positive boost $k$
used in our toy models makes the particles, initialy separated by average
distance $2 \sigma$, first approach each other and reach the interaction zone,
separation occurs a little later. The separation degree of freedom is the
relative variable, $x=x_1-x_2$, naturally, and it is natural to estimate the
separation by means of the averages,
$\langle x \rangle=\langle \psi_{rel} | x ] \psi_{rel} \rangle$ and
$\bar x=\langle \psi_{rel} | x^2 ] \psi_{rel} \rangle^{\frac{1}{2}}$,
the former being driven by the velocty, $2 k$, and the latter
taking spreading into account. As soon as one of such averages reaches, say,
one-and-a-half the range of $V$, the collision can be considered as finished
and $\rho_{12}$ can be considered as driven by $p_1^2/2+p_2^2/2$ only..
This suggests that $\bar x$ is of first order in $t$ when $t$ grows infinite.
The numerical estimates obtained from our various toy models confirm such a
linear trend, see for example Figure \ref{figxbar}, which shows linear trends
for all our toy cases.

There is no doubt that $V$ experiences a switch off, and, therefore, that the
various density operators of interest reach constant asymptotic entropies,
defining an asymptotic ${\cal I}$.

\section{VI Summary and Conclusion}
This work considers the non relativistic collision of two spinless particles.
Different toy models allow the calculation of time dependent wave functions
and density matrices, hence the calculation of the corresponding entropies.

Our numerical evidence is that the entanglement entropy $\cal{I}$, as a function
of time, increases monotonically and saturates, in a sigmo\" \i d pattern.
This answers the question left open by the authors of \cite{robi} in their
relativistic approach. We may conjecture that this sigmo\" \i d behavior holds
for all collisions, but the necessary theorem remains to be proved.

A further result is the confirmation of \cite{robi} in that the orders of
magnitude that we find for $\cal I$ in the non relativistic approach are the
same as those found in the relativistic one. Here again, however, a
general theorem providing bounds for $\cal{I}$, if they exist, is missing.

\section{Appendix: Entropy theorems}
We recall here two simple theorems

Th 1: {\it Unitary transforms leave the entropy invariant.}

Let $\rho$ be any density operator. Its negentropy, $-S$,  reads,
\begin{equation}
Tr\rho\log\rho = Tr\left[\frac{d}{dn}e^{n\log\rho}\right]_{n=1}
= Tr\left[\frac{d}{dn}\rho^n\right]_{n=1}.
\end{equation}
If the Hamiltonian $H$ is hermitian, the evolution operator,
$U \equiv e^{-iHt}$, is unitary. More generally, let $U$ be any unitary
operator transforming $\rho$ into $\rho'=\, U \rho U^{\dagger}$. The negentropy
becomes
\begin{eqnarray}
Tr\rho'\log\rho'=Tr\left[\frac{d}{dn}\rho^{\prime\, n}\right]_{n=1} =
Tr\left[\frac{d}{dn} U \rho^n U^{\dagger} \right]_{n=1}  \nonumber \\
= Tr U \left[\frac{d}{dn} \rho^n \right]_{n=1}  U^{\dagger}
= Tr\left[\frac{d}{dn}\rho^n\right]_{n=1},
\end{eqnarray}
because of the cyclic invariance of the Trace operation. Hence, no change.

Another, trivial argument to state for this invariance is that the unitary
transform modifies the eigenstates of $\rho$ and leaves its spectrum intact.
For the sake of the argument, assume that this spectrum is made of a finite
or infinite discrete set of eigenvalues $\lambda_i$. Hence the entropy,
calculated as, $S=-\sum_i \lambda_i\log\lambda_i$, is obviously invariant if
this spectrum is invariant.

Th 2: {\it If the Hamiltonian is separable between particles 1 and 2,
the one-body densities evolve independently.}

In our models, $H$ remains separable concerning cm and rel motions even when
$V(r) \neq 0$. But we set now $V=0$ and $H=H_1+H_2$. The Hamiltonian $H_1$
can contain both a kinetic energy and an external potential operators. The same
holds for $H_2$. The two-particle evolution operator, $U_{12}=e^{-i H t}$, then
factorises as $U_{12}=U_1U_2$, with commuting factors $U_1=e^{-i H_1 t}$ and 
$U_2=e^{-i H_2 t}$. The two-body density evolves in time as,
$\rho'_{12}=U_1U_2 \rho_{12} U_2^{\dagger}U_1^{\dagger}$. Then, as a partial trace,
the one-body density $\rho'_1$ reads,
\begin{eqnarray}
\rho'_1 = Tr_2 U_1U_2 \rho_{12} U_2^{\dagger}U_1^{\dagger} = 
U_1\, (Tr_2 U_2 \rho_{12} U_2^{\dagger})\, U_1^{\dagger} \nonumber \\
= U_1\, (Tr_2 \rho_{12})\, U_1^{\dagger} = U_1 \rho_1 U_1^{\dagger},
\end{eqnarray}
where we have again used the cyclic invariance of the Trace operation,
to get rid of $U_2$.
A similar result clearly holds for the second density evolution,
$\rho'_2 = U_2 \rho_2 U_2^{\dagger}$.

\end{document}